\documentclass{pnastwo}

\usepackage{graphicx}

\usepackage{amssymb,amsfonts,amsmath}

\newcommand{\ra}{\rangle}
\newcommand{\la}{\langle}
\newcommand{\beq}{\begin{equation}}
\newcommand{\eeq}{\end{equation}}
\newcommand{\vnabla}{\mbox{\boldmath $\nabla$}}
\newcommand{\vrr}{\mathbf{r}}

\newcommand{\vQ}{\mathbf{Q}}
\newcommand{\vJJ}{\mathbf{J}}
\newcommand{\cvJ}{\mbox{\boldmath ${\cal J}$}}
\newcommand{\cvA}{\mbox{\boldmath ${\cal A}$}}
\newcommand{\vj}{\mathbf{j}}
\newcommand{\vxi}{\mbox{\boldmath $\xi$}}
\newcommand{\vecep}{\mbox{\boldmath $\epsilon$}}
\newcommand{\ecep}{\mbox{$\epsilon$}}
\newcommand{\vom}{\mbox{\boldmath $\omega$}}
\newcommand{\om}{\mbox{$\omega$}}
\newcommand{\vvep}{\mbox{\boldmath $\varepsilon$}}
\newcommand{\vE}{\mathbf{E}}
\newcommand{\vla}{\mbox{\boldmath $\lambda$}}
\newcommand{\vnu}{\mbox{\boldmath $\nu$}}

\contributor{Submitted to Proceedings
of the National Academy of Sciences of the United States of America}
\url{www.pnas.org/cgi/doi/10.1073/pnas.0709640104}
\copyrightyear{2008}
\issuedate{Issue Date}
\volume{Volume}
\issuenumber{Issue Number}

\begin{document}



\title{Symmetries in Fluctuations Far from Equilibrium}





\author{P.I. Hurtado\affil{1}{Departamento de Electromagnetismo y F\'{\i}sica de la Materia, and 
Instituto Carlos I de F\'{\i}sica Te\'orica y Computacional, Universidad de Granada, 
Granada 18071, Spain},
C. P\'erez-Espigares\affil{1}{},
J.J. del Pozo\affil{1}{},
\and
P.L. Garrido\affil{1}{}%
}

\contributor{Submitted to Proceedings of the National Academy of Sciences
of the United States of America}

\maketitle

\begin{article}

\begin{abstract}
Fluctuations arise universally in Nature as a reflection of the discrete microscopic world at the macroscopic level. Despite their apparent noisy origin, 
fluctuations encode fundamental aspects of the physics of the system at hand, crucial to understand irreversibility and nonequilibrium behavior.
In order to sustain a given fluctuation, a system traverses a precise optimal path in phase space. Here we show that by demanding 
invariance of optimal paths under symmetry transformations, new and general fluctuation relations valid arbitrarily far from equilibrium are unveiled.
This opens an unexplored route toward a deeper understanding of nonequilibrium physics by bringing symmetry principles to the realm of 
fluctuations. We illustrate this concept studying symmetries of the current distribution out of equilibrium.
In particular we derive an isometric fluctuation relation which links in a strikingly simple manner the probabilities of any pair of isometric 
current fluctuations. This relation, which results from the time-reversibility of the dynamics, includes as a particular instance 
the Gallavotti-Cohen fluctuation theorem in this context but adds a completely new perspective on the high level of symmetry imposed by 
time-reversibility on the statistics of nonequilibrium fluctuations. 
The new symmetry implies remarkable hierarchies of equations for the current 
cumulants and the nonlinear response coefficients, going far beyond Onsager's reciprocity relations and Green-Kubo formulae. 
We confirm the validity of the new symmetry relation in extensive numerical 
simulations, and suggest that the idea of symmetry in fluctuations as invariance of optimal paths has far-reaching consequences in diverse fields.
\end{abstract}






\dropcap{L}arge fluctuations, though rare, play an important role in many fields of science as they crucially determine the fate of a system \cite{Ritort1}. 
Examples range from chemical reaction kinetics or the escape of metastable electrons in nanoelectronic devices to conformational 
changes in proteins, mutations in DNA, and nucleation events in the primordial universe. Remarkably, the statistics of these large 
fluctuations contains deep information on the physics of the system of interest \cite{BertiniC,Derrida}. 
This is particularly important for systems far from equilibrium, where 
no general theory exists up to date capable of predicting macroscopic and fluctuating behavior in terms of microscopic physics, 
in a way similar to equilibrium statistical physics. The consensus is that the study of fluctuations out of equilibrium may open the door to such general theory.
As most nonequilibrium systems are characterized by currents of locally conserved observables, 
understanding current statistics in terms of microscopic dynamics has become one of the main objectives of nonequilibrium statistical physics \cite{BertiniC,Derrida,BertiniA,BertiniB,BD,PabloA,PabloB,BD2,Livi,DharA,weA,weB,Evans,GC,LS,Kurchan}.
Pursuing this line of research is both of fundamental as well as practical importance. At the theoretical level, the function controlling current fluctuations
can be identified as the nonequilibrium analog of the free energy functional in equilibrium systems \cite{BertiniC,Derrida,BertiniA,BertiniB},
from which macroscopic properties of a nonequilibrium system can be obtained (including its most prominent features, as for instance the 
ubiquitous long range correlations \cite{Pedro,correl}, etc.) On the other hand, the physics of most modern mesoscopic 
devices is characterized by large fluctuations which determine their behavior and function.
In this way understanding current statistics in these systems is of great practical significance.

Despite the considerable interest and efforts on these issues, exact and general results valid arbitrarily far from equilibrium are still very scarce. 
The reason is that, while in equilibrium phenomena dynamics is irrelevant and the Gibbs distribution provides all the necessary information, 
in nonequilibrium physics dynamics plays a dominant role, even in the simplest situation of a nonequilibrium steady state \cite{BertiniC,Derrida,BertiniA,BertiniB}.
However, there is a remarkable exception to this absence of general results which has triggered an important surge in activity
since its formulation in the mid nineties. This is the fluctuation theorem, first discussed in the context of 
simulations of sheared fluids \cite{Evans}, and formulated rigorously by Gallavotti and Cohen under very general assumptions \cite{GC}. 
This theorem, which implies a relation between the probabilities of a given current fluctuation and the inverse event,
is a deep statement on the subtle consequences of time-reversal symmetry of microscopic dynamics at the macroscopic, irreversible level.
Particularly important here is the observation that symmetries are reflected at the fluctuating macroscopic level arbitrarily far from equilibrium.
Inspired by this illuminating result, we explore in this paper the behavior of the current distribution under symmetry transformations \cite{Gross}.
Key to our analysis is the observation that, in order to facilitate a given current fluctuation, the system traverses a well-defined optimal path 
in phase space \cite{BertiniC,Derrida,BertiniA,BertiniB,PabloA,PabloB,naturepath}. This path is, under very general conditions, invariant under certain 
symmetry transformations on the current. Using this invariance we show that for $d$-dimensional, time-reversible systems described by a 
locally-conserved field and possibly subject to a boundary-induced gradient and an external field $\vE$, the probability $\text{P}_{\tau}(\vJJ)$ 
of observing a current $\vJJ$ averaged over a long time $\tau$ obeys an \emph{isometric} fluctuation relation (IFR)
\beq 
\lim_{\tau\to \infty}\, \frac{1}{\tau} \, \ln \left[\frac{\displaystyle \text{P}_{\tau}(\vJJ)}{\displaystyle \text{P}_{\tau}(\vJJ')}\right] = 
\vecep \cdot (\vJJ-\vJJ') \, ,
\label{IFR}
\eeq
for any pair of isometric current vectors, $|\vJJ|=|\vJJ'|$. Here $\vecep=\vvep + \vE$ is a constant vector directly related to the rate of 
entropy production in the system, which depends on the boundary baths via $\vvep$ (see below).
\begin{figure}
\centerline{
\includegraphics[width=9cm]{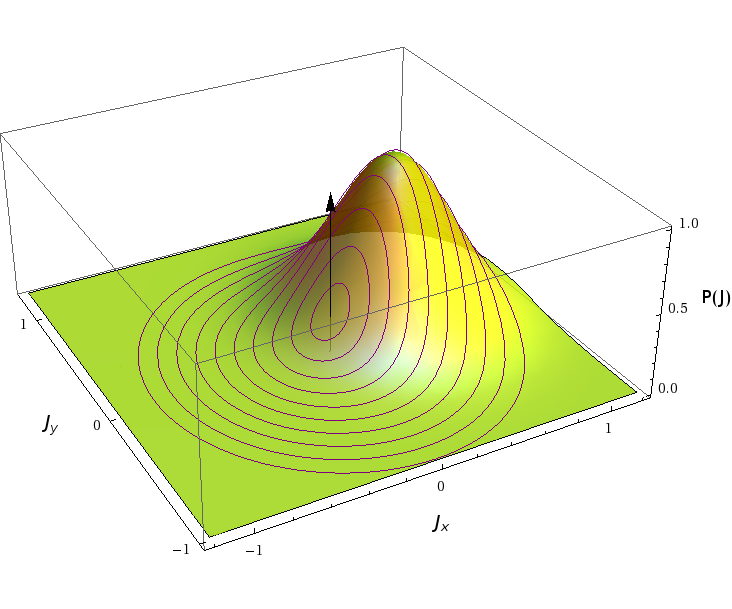}}
\caption{{The isometric fluctuation relation at a glance.} Sketch of the current distribution in two dimensions, peaked around its average 
$\la \vJJ\ra_{\epsilon}$, and isometric contour lines for different $|\vJJ|$'s. The isometric fluctuation relation, eq. (\ref{IFR}), establishes a simple relation 
for the probability of current fluctuations along each of these contour lines.}
\label{sketch}
\end{figure}

The above equation, which includes as a particular case the Gallavotti-Cohen (GC) result for $\vJJ'=-\vJJ$, relates in a strikingly simple manner
the probability of a given fluctuation $\vJJ$ with the likelihood of any other current fluctuation on the 
$d$-dimensional hypersphere of radius $|\vJJ|$, see Fig. 1, projecting a complex $d$-dimensional problem onto a much simpler 
one-dimensional theory. Unlike the GC relation which is a non-differentiable symmetry involving the inversion of the current sign, $\vJJ \to -\vJJ$,
eq. (\ref{IFR}) is valid for arbitrary changes in orientation of the current vector. This makes the experimental test of the above relation
a feasible problem, as data for current fluctuations involving different orientations around the average
can be gathered with enough statistics to ensure experimental accuracy. It is also important to notice that the isometric fluctuation relation is 
valid for arbitrarily large fluctuations, i.e. even for the non-Gaussian far tails of current distribution.
We confirm here the validity of the new symmetry in extensive numerical simulations of two different nonequilibrium systems: 
(i) A simple and very general lattice model of energy diffusion \cite{PabloA,PabloB,kmp},
and (ii) a hard-disk fluid in a temperature gradient \cite{GG}.

\subsection{The Isometric Fluctuation Relation}

Our starting point is a continuity equation which describes the macroscopic evolution of a wide class of systems characterized by a 
locally-conserved magnitude (e.g. energy, particle density, momentum, etc.)
\beq
\partial_t \rho(\vrr,t) = -\vnabla \cdot \Big( \vQ_{\vE}[\rho(\vrr,t)] + \vxi(\vrr,t) \Big) \, .
\label{langevin}
\eeq
Here $\rho(\vrr,t)$ is the density field, $\vj (\vrr,t)\equiv \vQ_{\vE}[\rho(\vrr,t)] + \vxi(\vrr,t)$ is the fluctuating current, 
with local average $\vQ_{\vE}[\rho(\vrr,t)]$, and $\vxi(\vrr,t)$ is a Gaussian white noise characterized by a 
variance (or \emph{mobility}) $\sigma[\rho(\vrr,t)]$. This (conserved) noise term accounts for microscopic random fluctuations at the macroscopic level.
Notice that the current functional includes in general the effect of a conservative external field, $\vQ_{\vE}[\rho(\vrr,t)]=\vQ[\rho(\vrr,t)] + \sigma[\rho(\vrr,t)] \vE$.
Examples of systems described by eq. (\ref{langevin}) range from diffusive systems  \cite{BertiniC,Derrida,BertiniA,BertiniB,BD,PabloA,PabloB,BD2},
where $\vQ[\rho(\vrr,t)]$ is given by Fourier's (or equivalently Fick's) law, $\vQ[\rho(\vrr,t)]=-D[\rho] \vnabla \rho(\vrr,t)$, to most interacting-particle 
fluids \cite{Spohn,Newman}, characterized by a Ginzburg-Landau-type theory for the locally-conserved particle density. To completely define the 
problem, the above evolution equation must be supplemented with appropriate boundary conditions, which may include an external gradient.

We are interested in the probability $\text{P}_{\tau}(\vJJ)$ of observing a space- and time-averaged empirical current $\vJJ$, defined as
\beq
\vJJ = \frac{1}{\tau}  \int_0^{\tau} dt \int d\vrr \, \, \vj (\vrr,t) \, .
\label{empJ}
\eeq
This probability obeys a large deviation principle for long times \cite{LDA,LDB},
$\text{P}_{\tau}(\vJJ)\sim \exp[+\tau L^d G(\vJJ)]$, where $L$ is the system linear size and $G(\vJJ)\le 0$ is the current large-deviation function (LDF), 
meaning that current fluctuations away from the average are exponentially unlikely in time. 
According to hydrodynamic fluctuation theory \cite{BertiniC,BertiniA,BertiniB,BD},
\beq
G(\vJJ) =-\min_{\rho(\vrr)} \int \frac{\displaystyle \left(\vJJ - \vQ_{\vE}[\rho(\vrr)]\right)^2} {\displaystyle 2 \sigma[\rho(\vrr)] } d \vrr \, ,
\label{ldf}
\eeq
which expresses the \emph{locally}-Gaussian nature of fluctuations \cite{BD,PabloA,PabloB}. 
The optimal profile $\rho_0(\vrr;\vJJ)$ solution of the above variational problem
can be interpreted as the density profile the system adopts to facilitate a current fluctuation $\vJJ$ \cite{PabloA,PabloB,naturepath}.
To derive eq. (\ref{ldf}) we assumed that
(\emph{i}) the optimal profiles associated to a given current fluctuation 
are time-independent \cite{BertiniC,Derrida,BertiniA,BertiniB,BD,PabloA,PabloB,BD2,naturepath,Pablo3}, 
and (\emph{ii}) the optimal current field has no spatial structure, see Supporting Information (SI).
This last hypothesis, which greatly simplifies the calculation of current statistics, can be however relaxed for our purposes (as shown below).
The probability $\text{P}_{\tau}(\vJJ)$ is thus simply the \emph{Gaussian} weight associated to the optimal profile.
Note however that the minimization procedure gives rise to a nonlinear problem which 
results in general in a current distribution with non-Gaussian tails \cite{BertiniC,Derrida,BertiniA,BertiniB,BD,PabloA,PabloB}.

The optimal profile is solution of the following equation
\beq
\frac{\delta \om_2[\rho(\vrr)]}{\delta \rho(\vrr')} - 2\vJJ\cdot \frac{\delta \vom_1[\rho(\vrr)]}{\delta \rho(\vrr')} + 
\vJJ^2 \frac{\delta \om_0[\rho(\vrr)]}{\delta \rho(\vrr')}  = 0 \, ,
\label{optprof}
\eeq
where $\frac{\delta}{\delta \rho(\vrr')}$ stands for functional derivative, and
\beq
\vom_n[\rho(\vrr)] \equiv  \int d\vrr \frac{\vQ_{\vE}^n[\rho(\vrr)]}{\sigma[\rho(\vrr)]} \, . 
\label{defs}
\eeq
Remarkably, the optimal profile $\rho_0(\vrr;\vJJ)$ solution of eq. (\ref{optprof}) depends exclusively on $\vJJ$ and $\vJJ^2$.
Such a simple quadratic dependence, inherited from the locally-Gaussian nature of fluctuations,
has important consequences at the level of symmetries of the current distribution. In fact, it is clear from eq. (\ref{optprof}) that the condition
\beq
\frac{\delta \vom_1[\rho(\vrr)]}{\delta \rho(\vrr')} = 0 \, ,
\label{cond1}
\eeq
implies that $\rho_0(\vrr;\vJJ)$ will depend exclusively on the \emph{magnitude} of the current vector, 
via $\vJJ^2$, not on its \emph{orientation}. In this way, all isometric current fluctuations characterized by a constant $|\vJJ|$ will
have the same associated optimal profile, $\rho_0(\vrr;\vJJ)=\rho_0(\vrr;|\vJJ|)$, independently of whether the 
current vector $\vJJ$ points along the gradient direction, against it, or along any arbitrary direction. In other words, the optimal profile
is invariant under current rotations if eq. (\ref{cond1}) holds. It turns out that condition (\ref{cond1}) follows from the time-reversibility of the dynamics, 
in the sense that the evolution operator in the Fokker-Planck formulation of eq. (\ref{langevin}) obeys a local detailed balance condition \cite{LS,Kurchan}. 
In this case $\vQ_{\vE}[\rho(\vrr)]/\sigma[\rho(\vrr)] = -\vnabla \delta {\cal H}[\rho]/\delta\rho$, with ${\cal H}[\rho(\vrr)]$ the system Hamiltonian, and condition (\ref{cond1}) holds.
The invariance of the optimal profile can be now used in eq. (\ref{ldf}) to relate in a simple way
the current LDF of any pair of isometric current fluctuations $\vJJ$ and $\vJJ'$, with $|\vJJ|=|\vJJ'|$,
\beq
G(\vJJ) - G(\vJJ') = |\vecep| |\vJJ| (\cos \theta - \cos \theta') \, ,
\label{IFRcos}
\eeq
where $\theta$ and $\theta'$ are the angles formed by vectors $\vJJ$ and $\vJJ'$, respectively, with a constant vector 
$\vecep=\vvep + \vE$, see below. Eq. (\ref{IFRcos}) is just an alternative formulation of the isometric fluctuation relation (\ref{IFR}). 
By letting $\vJJ$ and $\vJJ'$ differ by an infinitesimal angle, the IFR can be cast in a simple differential 
form, $\partial_{\theta} G(\vJJ) = |\vecep| |\vJJ| \sin \theta$, which reflects the high level of symmetry imposed by time-reversibility on the 
current distribution.

The condition $\delta \vom_1[\rho(\vrr)]/\delta \rho(\vrr')=0$ can be seen as a conservation law. It implies that the observable $\vom_1[\rho(\vrr)]$
is in fact a \emph{constant of motion}, $\vecep\equiv \vom_1[\rho(\vrr)]$, independent of the profile $\rho(\vrr)$,
which can be related with the rate of entropy 
production via the Gallavotti-Cohen theorem \cite{GC,LS,Kurchan}. In a way similar to Noether's theorem, the conservation law for $\vecep$ implies a symmetry
for the optimal profiles under rotations of the current and a fluctuation relation for the current LDF. This constant can 
be easily computed under very general assumptions (see SI).

\subsection{Implications and Generalizations}

The isometric fluctuation relation, eq. (\ref{IFR}), has far-reaching and nontrivial consequences. First, 
the IFR implies a remarkable hierarchy of equations for the cumulants of the current distribution, see eq. (\ref{cumul}) in Methods.
This hierarchy can be derived starting from the Legendre transform of the current LDF, $\mu(\vla)=\max_{\vJJ}[G(\vJJ)+\vla\cdot \vJJ]$, from which
all cumulants can be obtained \cite{Derrida}, and writing the IFR for $\mu(\vla)$ in the limit of infinitesimal rotations.
As an example, the cumulant hierarchy in two dimensions implies the following relations
\begin{eqnarray}
 \la J_x\ra _{\epsilon}& = & \tau L^2 \left[\epsilon_x \la\Delta J_y^2\ra _{\epsilon}- \epsilon_y\la\Delta J_x \Delta J_y\ra_{\epsilon}  \right] \, \label{cumul2dn1} \\
 \la J_y\ra_{\epsilon} & = & \tau L^2 \left[\epsilon_y \la\Delta J_x^2\ra_{\epsilon} - \epsilon_x\la\Delta J_x \Delta J_y\ra_{\epsilon}  \right] \, \nonumber \\
\phantom{aaa} \nonumber \\
2\la \Delta J_x \Delta J_y\ra_{\epsilon} & = & \tau L^2 \left[\epsilon_y \la\Delta J_x^3\ra_{\epsilon} - \epsilon_x\la\Delta J_x^2 \Delta J_y\ra_{\epsilon}  \right]  \, \label{cumul2dn2} \\
& = & \tau L^2 \left[\epsilon_x \la\Delta J_y^3\ra_{\epsilon} - \epsilon_y\la\Delta J_x \Delta J_y^2\ra_{\epsilon}  \right] \, \nonumber \\
\la \Delta J_x^2\ra_{\epsilon} - \la \Delta J_y^2\ra_{\epsilon} & = & \tau L^2 \left[\epsilon_x \la\Delta J_x \Delta J_y^2\ra_{\epsilon} - \epsilon_y\la\Delta J_x^2 \Delta J_y\ra_{\epsilon}  \right] \, , \nonumber
\end{eqnarray}
for the first cumulants, with $\Delta J_{\alpha}\equiv J_{\alpha}-\la J_{\alpha}\ra_{\epsilon}$. 
It is worth stressing that the cumulant hierarchy is valid arbitrarily far from equilibrium. 
In a similar way,  the IFR implies a set of hierarchies for the nonlinear response coefficients, see eqs. (\ref{nonlincoefs1})-(\ref{nonlincoefs3}) in Methods. 
In our two-dimensional example, let $_{(n)}^{(k)}\chi_{(n_x,n_y)}^{(k_x,k_y)}$ be the response coefficient of the 
cumulant $\la\Delta J_x^{n_x} \Delta J_y^{n_y}\ra_{\epsilon}$ 
to order $\ecep_x^{k_x}\ecep_y^{k_y}$, with $n=n_x+n_y$ and $k=k_x+k_y$. 
To the lowest order these hierarchies imply Onsager's reciprocity symmetries and Green-Kubo relations for the linear response coefficients of
the current. They further predict that in fact the linear response matrix is proportional to the identity, so 
$_{(1)}^{(1)}\chi_{(1,0)}^{(1,0)}={_{(1)}^{(1)}\chi_{(0,1)}^{(0,1)}}={_{(2)}^{(0)}\chi_{(2,0)}^{(0,0)}}={_{(2)}^{(0)}\chi_{(0,2)}^{(0,0)}}$ while $_{(1)}^{(1)}\chi_{(1,0)}^{(0,1)}=0={_{(1)}^{(1)}\chi_{(0,1)}^{(1,0)}}$.
The first nonlinear coefficients of the current can be simply written in terms of the linear coefficients of the second cumulants as
$_{(1)}^{(2)}\chi_{(1,0)}^{(2,0)} = 2 {_{(2)}^{(1)}\chi_{(2,0)}^{(1,0)}}$ and $_{(1)}^{(2)}\chi_{(1,0)}^{(0,2)} = -2 {_{(2)}^{(1)}\chi_{(1,1)}^{(1,0)}}$, 
while the cross-coefficient reads $_{(1)}^{(2)}\chi_{(1,0)}^{(1,1)} = 2 [{_{(2)}^{(1)}\chi_{(2,0)}^{(0,1)}} + {_{(2)}^{(1)}\chi_{(1,1)}^{(0,1)}}]$
(symmetric results hold for $n_x=0$, $n_y=1$). Linear response coefficients for the second-order cumulants also obey simple relations,
e.g. $_{(2)}^{(1)}\chi_{(1,1)}^{(1,0)} = - {_{(2)}^{(1)}\chi_{(1,1)}^{(0,1)}}$ and 
$_{(2)}^{(1)}\chi_{(2,0)}^{(1,0)} + {_{(2)}^{(1)}\chi_{(2,0)}^{(0,1)}} = {_{(2)}^{(1)}\chi_{(0,2)}^{(1,0)}} + {_{(2)}^{(1)}\chi_{(0,2)}^{(0,1)}}$, and 
the set of relations continues to arbitrary high orders. In this way hierarchies (\ref{nonlincoefs1})-(\ref{nonlincoefs3}),
which derive from microreversibility as reflected in the IFR, provide deep insights into nonlinear response theory for nonequilibrium systems \cite{Gaspard}.

The IFR and the above hierarchies all follow from the invariance of optimal profiles under certain transformations. This idea can be further exploited 
in more general settings. In fact, by writing explicitly the dependence on the external field $\vE$ in eq. (\ref{optprof}) for the optimal profile, one realizes 
that if $\frac{\delta}{\delta \rho(\vrr')} \int \vQ[\rho(\vrr)] d\vrr = 0$, 
together with the time-reversibility condition, eq. (\ref{cond1}), the resulting optimal profiles are invariant under \emph{independent} rotations of the current and
the external field. It thus follows that the current LDFs for pairs $(\vJJ,\vE)$ and $(\vJJ'={\cal R} \vJJ,\vE^*={\cal S} \vE)$, 
with ${\cal R}$, ${\cal S}$ independent rotations, obey a generalized
isometric fluctuation relation
\beq
G_{\vE}(\vJJ) - G_{\vE^*}(\vJJ') = \vvep\cdot(\vJJ-\vJJ') - \vnu\cdot(\vE-\vE^*) + \vJJ\cdot\vE-\vJJ'\cdot\vE^* \, ,
\label{IFR2}
\eeq
where we write explicitly the dependence of the current LDF on the external field. The vector $\vnu\equiv \int \vQ[\rho(\vrr)] d\vrr$ is now another constant 
of motion, independent of $\rho(\vrr)$, which can be easily computed (see SI). For a fixed boundary gradient, 
the above equation relates any current fluctuation $\vJJ$ in the 
presence of an external field $\vE$ with any other isometric current fluctuation $\vJJ'$ in the presence of an arbitrarily-rotated external field 
$\vE^*$, and reduces to the standard IFR for $\vE=\vE^*$. Condition $\frac{\delta}{\delta \rho(\vrr')} \int \vQ[\rho(\vrr)] d\vrr = 0$ is rather general, as most time-reversible systems with a 
local mobility $\sigma[\rho]$ do fulfill this condition (e.g., diffusive systems).

The IFR can be further generalized to cases where the current profile is not constant, relaxing hypothesis (\emph{ii}) above. Let $\text{P}_{\tau}[\cvJ(\vrr)]$ be the probability of 
observing a time-averaged 
current field $\cvJ(\vrr)=\tau^{-1}\int_0^{\tau} dt \, \vj (\vrr,t)$. This vector field must have zero divergence because it is coupled via the continuity equation to an optimal density profile which is 
assumed to be time-independent, see SI and hypothesis (\emph{i}) above. 
Because of time-reversibility, $\vQ_{\vE}[\rho(\vrr)]/\sigma[\rho(\vrr)] = -\vnabla \delta {\cal H}[\rho]/\delta\rho$ and it is easy to show in the equation for the optimal density profile that the term linear in 
$\cvJ(\vrr)$ vanishes, so $\rho_0[\vrr;\cvJ(\vrr)]$ remains invariant under (local or global) rotations of $\cvJ(\vrr)$, see SI. In this way, for any divergence-free current field $\cvJ'(\vrr)$ locally-isometric to 
$\cvJ(\vrr)$, so $\cvJ'(\vrr)^2 = \cvJ(\vrr)^2$ $\forall \vrr$, we can write a generalized isometric fluctuation relation
\beq
\lim_{\tau\to \infty}\, \frac{1}{\tau} \, \ln \left[\frac{\displaystyle \text{P}_{\tau}[\cvJ(\vrr)]}{\displaystyle \text{P}_{\tau}[\cvJ'(\vrr)]}\right] = 
\int_{\partial\Lambda} d\Gamma \, \frac{\delta {\cal H}[\rho]}{\delta \rho} \hat{n} \cdot [\cvJ'(\vrr)-\cvJ(\vrr)] \, ,
\label{gIFR}
\eeq
where the integral (whose result is independent of $\rho(\vrr)$) is taken over the boundary $\partial \Lambda$ of the domain $\Lambda$ where the system is defined, and $\hat{n}$ is the unit vector 
normal to the boundary at each point. Eq. (\ref{gIFR}) generalizes the IFR to situations where hypothesis (\emph{ii}) is violated, opening the door to isometries based on local (in addition to global) rotations.
As a corollary, we show in the SI appendix that a similar generalization of the isometric fluctuation symmetry does not exist whenever optimal profiles become time-dependent,
so the IFR breaks down in the regime where hypothesis (\emph{i}) is violated.
In this way, we may use violations of the IFR and its generalizations to detect the instabilities which characterize the fluctuating behavior of the system at hand \cite{BertiniC,BD2,Pablo3}.

\subsection{Checking the Isometric Fluctuation Relation}

We have tested the validity of the IFR in extensive numerical simulations of two different nonequilibrium systems. The first one is a simple and very 
general model of energy diffusion \cite{PabloA,PabloB,kmp} defined on a two-dimensional (2D) square lattice with $L^2$ sites. Each site is characterized
by an energy $e_i$, $i\in[1,L^2]$, and models a harmonic oscillator which is mechanically uncoupled from its nearest neighbors but interact with them 
via a stochastic energy-redistribution process. Dynamics thus proceeds through random energy exchanges between randomly-chosen nearest 
neighbors. In addition, left and right boundary sites may interchange energy with boundary baths at temperatures $T_L$ and $T_R$, respectively, while
periodic boundary conditions hold in the vertical direction. For $T_L\neq T_R$ the systems reaches a nonequilibrium steady state characterized, in the absence of 
external field (the case studied here), by a linear energy profile $\rho_{\text{st}}(\vrr)=T_L + x\,(T_R-T_L)$ and a nonzero average current given by 
Fourier's law. This model plays a fundamental role in nonequilibrium statistical physics as 
a testbed to assess new theoretical advances, and represents at a coarse-grained level a large class of diffusive systems of technological and 
theoretical interest \cite{PabloA,PabloB}. The model is described at the macroscopic level by eq. (\ref{langevin}) with a diffusive current term 
$\vQ[\rho(\vrr,t)]=-D[\rho]\vnabla\rho$ with $D[\rho]=\frac{1}{2}$ and $\sigma[\rho]=\rho^2$, and it turns out to be an optimal candidate to test the 
IFR because: (1) the associated hydrodynamic fluctuation theory can be solved analytically \cite{Carlos}, and (2) its dynamics is simple enough to 
allow for a detailed numerical study of current fluctuations.

In order to test the IFR in this model we performed a large number of steady-state simulations of long duration $\tau>L^2$ (the unit of time is the 
Monte Carlo step) for $L=20$, $T_L=2$ and $T_R=1$, accumulating statistics for the space- and time-averaged current vector $\vJJ$. 
The measured current distribution is shown in the bottom inset to Fig. 2, together with a fine polar binning which allows us to compare the 
probabilities of isometric current fluctuations along each polar corona, see eq. (\ref{IFR}). Taking $G(\vJJ)=(\tau L^d)^{-1} \ln \text{P}_{\tau}(\vJJ)$, 
Fig. 2 confirms the IFR prediction that $G(\vJJ)-G(\vJJ')$, once scaled by $|\vJJ|^{-1}$, collapses onto a linear function of $\cos\theta-\cos\theta'$ 
for all values of $|\vJJ|$, see eq. (\ref{IFRcos}). Here $\theta$, $\theta'$ are the angles formed by the isometric current vectors $\vJJ$, $\vJJ'$ with the $x$-axis 
($\vE=0$ in our case). We also measured the average energy profile associated to each current fluctuation, $\rho_0(\vrr;\vJJ)$, see top inset to 
Fig. 2.  As predicted above, profiles for different but isometric current fluctuations 
all collapse onto a single curve, confirming the invariance of optimal profiles under current rotations. 

Standard simulations allow us to explore moderate fluctuations of the current around the average. In order to test the IFR in the far tails of the 
current distribution, corresponding to exponentially unlikely rare events, we implemented an elegant method recently introduced to measure large 
deviation functions in many-particle systems \cite{sim}. The method, which yields the Legendre transform of the current LDF, $\mu(\vla)$, is based 
on a modification of the dynamics so that the rare events responsible of the large deviation are no longer rare \cite{sim}, and has been recently used 
with success to confirm an additivity conjecture regarding large fluctuations in nonequilibrium systems \cite{PabloA,PabloB}.
Using this method we measured $\mu(\vla)$ in increasing manifolds of constant $|\vla+\vecep|$, see Fig. 3. The IFR implies that $\mu(\vla)$ is constant
along each of these manifolds, or equivalently $\mu(\vla)=\mu[{\cal R}_{\phi}(\vla+\vecep)-\vecep]$, $\forall\phi\in[0,2\pi]$, with ${\cal R}_{\phi}$ 
a rotation in 2D of angle $\phi$. Fig. 3 shows the measured $\mu(\vla)$ for different values of $|\vla+\vecep|$ 
corresponding to very large current fluctuations, different rotation angles $\phi$ and increasing system sizes, together with the theoretical 
predictions \cite{Carlos}. As a result of the finite, discrete character of the lattice system studied here, we observe weak violations of IFR in the far
tails of the current distribution, specially for
currents orthogonal to $\vecep$. These weak violations are expected since a prerequisite for the IFR to 
hold is the existence of a macroscopic limit, i.e. eq. (\ref{langevin}) should hold strictly, which is not the case for the relatively small values of $L$ studied here. 
However, as $L$ increases, a clear convergence toward the IFR prediction is observed as the effects associated to the underlying 
lattice fade away, strongly supporting the validity of IFR in the macroscopic limit.

We also measured current fluctuations in a Hamiltonian hard-disk fluid subject to a temperature gradient \cite{GG}. This model is a paradigm in 
liquid state theory, condensed matter and statistical physics, and has been widely studied during last decades. The model consists in $N$ hard disks
of unit diameter interacting via instantaneous collisions and confined to a box of linear size $L$ such that the particle density is fixed to $\Phi=N/L^2=0.58$. 
Here we choose $N=320$. The box is divided in three parts: a central, bulk region of width $L-2\alpha$ with periodic boundary conditions in the vertical 
direction, and two lateral stripes of width $\alpha=L/4$ which act as deterministic heat baths, see bottom inset to Fig. 4. 
This is achieved by keeping constant the total kinetic energy within each lateral band via a global, instantaneous rescaling of the velocity of bath 
particles after bath-bulk particle collisions. This heat bath mechanism has been shown to efficiently thermostat the fluid \cite{GG}, and 
has the important advantage of being deterministic. As for the previous diffusive model, we performed a large number of steady state simulations of 
long duration ($\tau> 2N$ collisions per particle) for $T_L=4$ and $T_R=1$, accumulating statistics for the 
current $\vJJ$ and measuring the average temperature profile associated to each $\vJJ$. Fig. 4 shows the linear collapse of $|\vJJ|^{-1}[G(\vJJ)-G(\vJJ')]$ 
as a function of $\cos\theta-\cos\theta'$ for different values of $|\vJJ|$, confirming the validity of the IFR for this hard-disk fluid in the moderate 
range of current fluctuations that we could access. Moreover, the measured optimal profiles for different isometric current fluctuations all nicely 
collapse onto single curves, see top inset to Fig. 4, confirming their rotational invariance. 

It is interesting to notice that the hard-disk fluid is a fully hydrodynamic system, with 4 different locally-conserved coupled fields possibly subject to memory effects, 
defining a far more complex situation than the one studied here, see eq. (\ref{langevin}). Therefore the validity of IFR in this context
suggests that this fluctuation relation, based on the invariance of optimal profiles under symmetry transformations, is in fact a rather general result valid
for arbitrary fluctuating hydrodynamic systems.
\begin{figure}[t!]
\vspace{-0.8cm}
\centerline{
\includegraphics[width=9.8cm,clip]{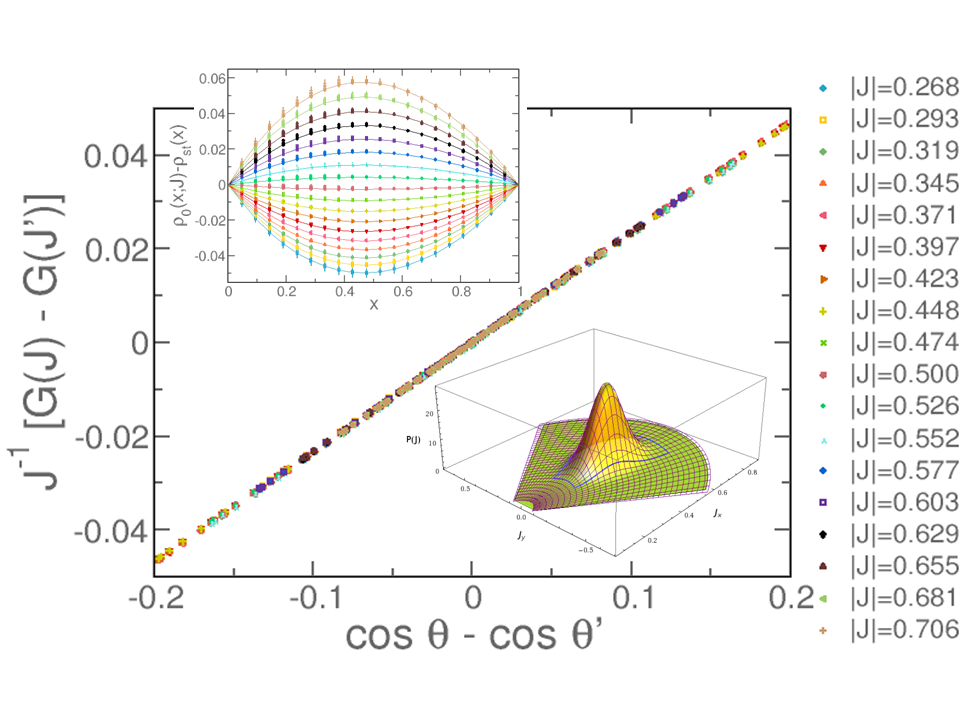}}
\caption{{Confirmation of IFR in a diffusive system.} The IFR predicts that $|\vJJ|^{-1}[G(\vJJ)-G(\vJJ')]$ collapses onto a linear function of $\cos\theta-\cos\theta'$ for all values of $|\vJJ|$. This collapse is confirmed here in the energy diffusion model for a wide range of values for $|\vJJ|$. 
Bottom inset: Measured current distribution together with the polar binning used to test the IFR. Top inset: Average profiles for different but isometric 
current fluctuations all collapse onto single curves, confirming the invariance of optimal profiles under current rotations. Angle range is 
$|\theta|\le 16.6^{\circ}$, see marked region in the histogram.
}
\label{kipnis1}
\end{figure}
\begin{figure}[t!]
\vspace{-0.8cm}
\centerline{
\includegraphics[width=9cm,clip]{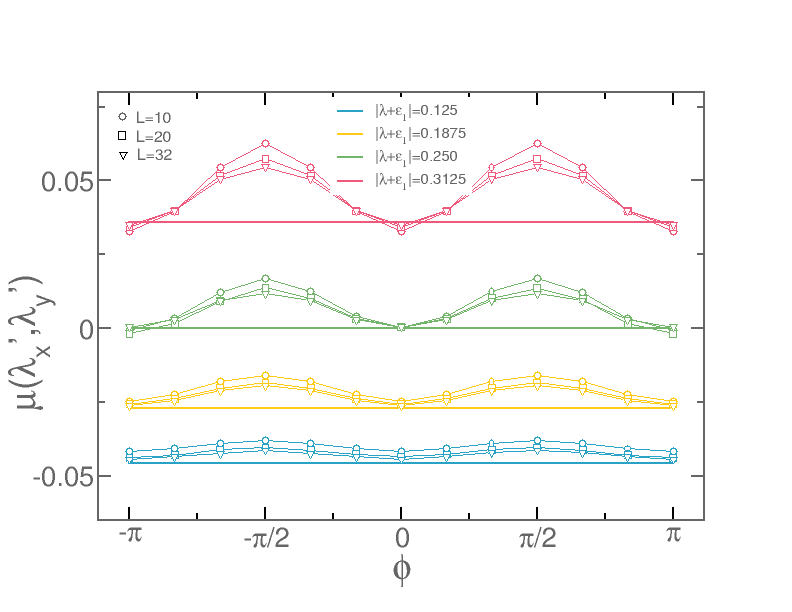}}
\caption{{IFR for large current fluctuations.} Legendre transform of the current LDF for the energy diffusion model, for different values of $|\vla+\vecep|$ 
corresponding to very large current fluctuations, different rotation angles $\phi$ such that $\vla'={\cal R}_{\phi}(\vla+\vecep)-\vecep$, and increasing system sizes. 
Lines are theoretical predictions. The IFR predicts that $\mu(\vla)=\mu[{\cal R}_{\phi}(\vla+\vecep)-\vecep]$ $\forall \phi\in[0,2\pi]$. The isometric fluctuation symmetry 
emerges in the macroscopic limit as the effects associated to the underlying lattice fade away.
}
\label{kipnis2}
\end{figure}

A few remarks are now in order. 
First, as a corollary to the IFR, it should be noted that for time-reversible systems with additive fluctuations, i.e. with a constant, profile-independent 
mobility $\sigma$, the optimal profile associated to a given current fluctuation is in fact independent of $\vJJ$, see eq. (\ref{optprof}), and hence 
equal to the stationary profile. In this case it is easy to show that current fluctuations are Gaussian, with 
$G(\vJJ)=\vecep\cdot(\vJJ-\la\vJJ\ra_{\epsilon}) + \sigma^{-1}(\vJJ^2-\la\vJJ\ra_{\epsilon}^2)$. This is the case, for instance, of model B in the 
Hohenberg-Halperin classification \cite{Newman} 
\footnote{Notice that $\rho$-dependent corrections to a constant mobility $\sigma$, which are typically irrelevant from a
renormalization-group point of view \cite{Newman}, turn out to be essential for current fluctuations as they give 
rise to non-Gaussian tails in the current distribution.}.
On the other hand, it should be noticed that the time-reversibility condition for the IFR to hold, eq. (\ref{cond1}) , is just a \emph{sufficient} but not 
necessary condition. In fact, we cannot discard the possibility of time-irreversible systems such that, only 
for the optimal profiles, $\delta \vom_1[\rho(\vrr)]/\delta \rho(\vrr')\vert_{\rho_0}=0$. 
\begin{figure}[t]
\vspace{-1.25cm}
\centerline{
\includegraphics[width=10.5cm,clip]{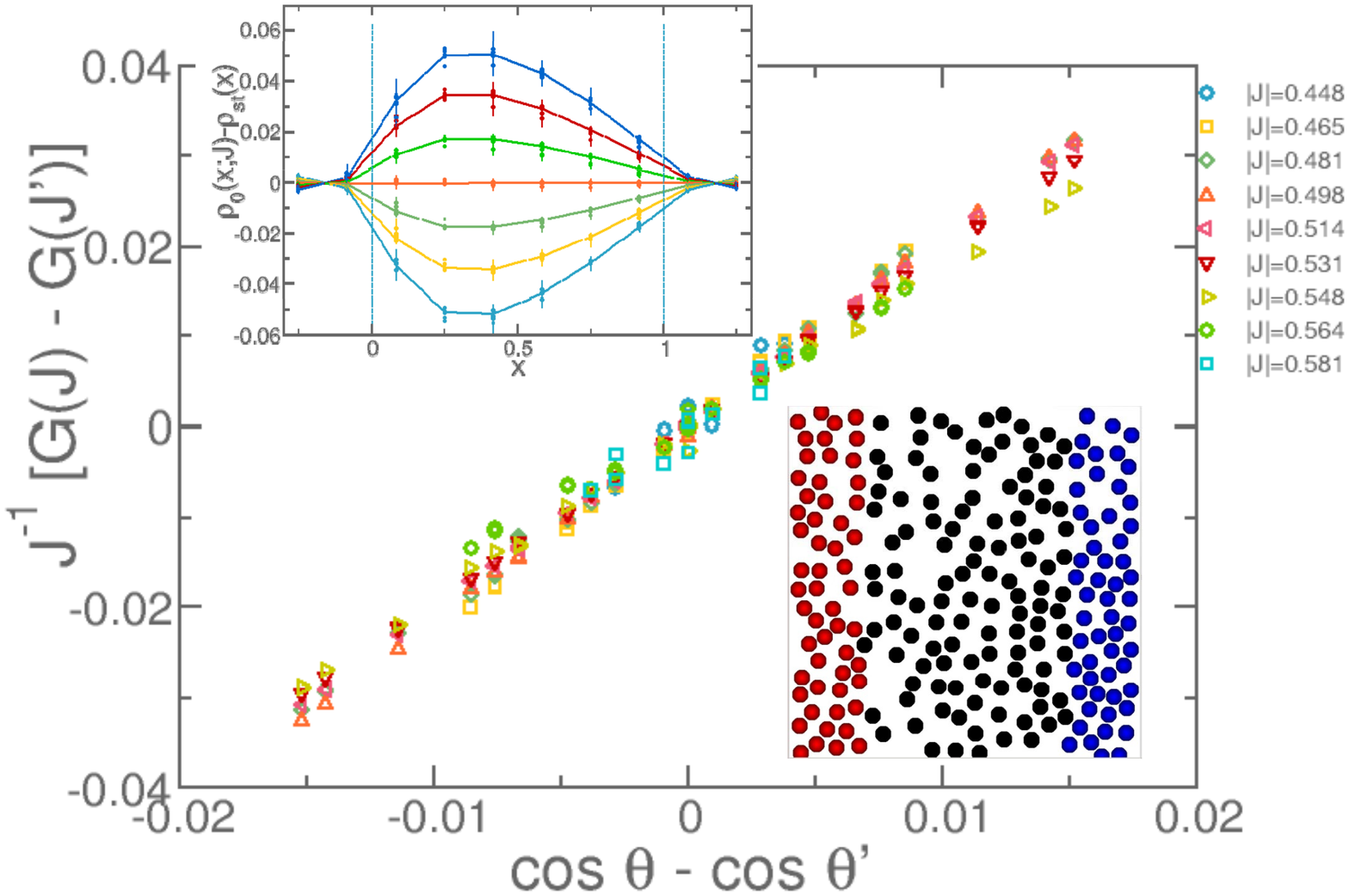}}
\vspace{-0.25cm}
\caption{{IFR in a hydrodynamic hard-disk fluid.} Confirmation of IFR in a two dimensional hard-disk fluid under a temperature gradient after a polar 
binning of the measured current distribution. As predicted by IFR, the difference of current LDFs for different isometric current fluctuations, once scaled by 
the current norm, collapses in a line when plotted against $\cos\theta-\cos\theta'$. Top inset: Optimal temperature profiles associated to different current fluctuations. Profiles for a given $|\vJJ|$ and different angles $\theta\in[-7.5^{\circ},+7.5^{\circ}]$ all collapse onto a single curve, thus confirming the 
invariance of optimal profiles under current rotations. Notice that the profiles smoothly penetrate into the heat baths. 
Bottom inset: Snapshot of the 2D hard-disk fluid with Gaussian heat baths.
}
\label{hd}
\end{figure}

\subsection{Discussion}

The IFR is a consequence of time-reversibility for systems in the hydrodynamic scaling limit, and reveals an unexpected high level of symmetry in the statistics of nonequilibrium fluctuations.
It generalizes and comprises the Gallavotti-Cohen fluctuation theorem for currents, relating the probabilities of an event not only with its time-reversal but with any other isometric fluctuation.
This has important consequences in the form of hierarchies for the current cumulants and the linear and nonlinear response coefficients, which hold arbitrarily far from equilibrium and can be readily tested in experiments.
A natural question thus concerns the level of generality of the isometric fluctuation relation. In this paper we have demonstrated the IFR for a broad class of systems characterized at the macroscale by a single conserved field,
using the tools of hydrodynamic fluctuation theory (HFT). This theoretical framework, summarized in the path large deviation functional, eq. (3) in the SI appendix, has been rigorously proven for a number of interacting particle 
systems \cite{BertiniC,Derrida,BertiniA,BertiniB}, but it is believed to remain valid for a much larger class of systems. The key is that the Gaussian nature of local fluctuations, which lies at the heart of the approach,
is expected to emerge for most situations in the appropriate macroscopic limit as a result of a central limit theorem: although microscopic interactions can be extremely complicated, the ensuing fluctuations 
of the slow hydrodynamic fields result from the sum of an enormous amount of random events at the microscale which give rise to Gaussian statistics. 
There exist of course anomalous systems for which local fluctuations at the macroscale can be non-Gaussian. In these cases we cannot discard that a modified version of the IFR could remain valid, though the analysis 
would be certainly more complicated.
Furthermore, our numerical results show that the IFR remains true even in cases where it is not clear whether the HFT applies, strongly supporting the validity of this symmetry for arbitrary fluctuating hydrodynamic systems.

A related question is the demonstration of the IFR starting from microscopic dynamics. Techniques similar to those in Refs. \cite{LS,Pablo2}, which derive the Gallavotti-Cohen fluctuation theorem from 
the spectral properties of the microscopic stochastic evolution operator, can prove useful for this task. However, in order to prove the IFR these techniques must be supplemented with additional insights on the asymptotic 
properties of the microscopic transition rates as the macroscopic limit is approached. In this way we expect finite-size corrections to the IFR which decay with the system size, as it is in fact observed in our simulations for 
the energy diffusion model, see Fig. 3. Also interesting is the possibility of an IFR for discrete isometries related with the underlying lattice in stochastic models. These open questions call for further study.

We have shown in this paper how symmetry principles come forth in fluctuations far from equilibrium.
By demanding invariance of the optimal path responsible of a given fluctuation under symmetry transformations, we unveiled a novel and very 
general isometric fluctuation relation for time-reversible systems  which relates in a simple manner the probability of any pair of isometric 
current fluctuations. Invariance principles of this kind can be applied with great generality in diverse fields where fluctuations play a fundamental role, 
opening the door to further exact and general results valid arbitrarily far from equilibrium. This is particularly relevant in mesoscopic biophysical
systems, where relations similar to the isometric fluctuation relation might be used to efficiently measure free-energy differences
in terms of work distributions \cite{Ritort}. Other interesting issues concern the study of general fluctuation relations emerging from the invariance 
of optimal paths in full hydrodynamical systems with several conserved fields, or the quantum analog of the isometric fluctuation relation in
full counting statistics.




\appendix[Hierarchies for the cumulants and response coefficients]

The moment-generating function associated to $\text{P}_{\tau}(\vJJ)$, defined as 
$\Pi_{\tau}(\vla)=\int \text{P}_{\tau}(\vJJ)\exp(\tau L^d \vla \cdot \vJJ)d\vJJ$, scales for long times as 
$\Pi_{\tau}(\vla)\sim \exp[+\tau L^d \mu(\vla)]$, where $\mu(\vla)=\max_{\vJJ}[G(\vJJ)+\vla\cdot \vJJ]$ is the Legendre transform of the current LDF.
The cumulants of the current distribution can be obtained from the derivatives of $\mu(\vla)$ evaluated at $\vla=0$, i.e. 
$\mu_{(n_1 ... n_d)}^{(n)}\equiv [\partial^n \mu(\vla)/\partial \lambda_1^{n_1} ... \lambda_d^{n_d}]_{\lambda=0} = (\tau L^d)^{n-1} \la\Delta J_1^{n_1}... \Delta J_d^{n_d} \ra_{\epsilon}$ for $n\ge 1$,
where $\Delta J_{\alpha}\equiv J_{\alpha}-(1-\delta_{n,1})\la J_{\alpha}\ra_{\epsilon}$ and $\delta_{n,m}$ is the Kronecker symbol. The IFR can be stated for the Legendre 
transform of the current LDF as $\mu(\vla)=\mu[{\cal R}(\vla+\vecep)-\vecep]$, where ${\cal R}$ is any $d$-dimensional rotation. 
Using this relation in the definition of the $n$-th order cumulant in the limit of infinitesimal rotations, ${\cal R}=\mathbb{I}+\Delta \theta {\cal L}$, it is easy to show that 
\beq
n_{\alpha} {\cal L}_{\beta \alpha} \mu_{(n_1 ... n_{\alpha}-1 ... n_{\beta}+1 ... n_d)}^{(n)} +   \epsilon_{\nu} {\cal L}_{\gamma\nu} \mu_{(n_1 ... n_{\gamma}+1 ... n_d)}^{(n+1)}= 0 \, ,
\label{cumul}
\eeq
where ${\cal L}$ is any generator of $d$-dimensional rotations, and summation over repeated Greek indices ($\in[1,d]$) is assumed. 
The above hierarchy relates in a simple way cumulants of orders $n$ and $n+1$ $\forall n \ge 1$, and is valid arbitrarily far from equilibrium. 
As an example, eqs. (\ref{cumul2dn1}) and (\ref{cumul2dn2}) above show the first two sets of relations ($n=1,2$) of the above hierarchy in two dimensions.
In a similar way, we can explore the consequences of the IFR on the linear and nonlinear response coefficients. For that, we now expand 
the cumulants of the current in powers of $\vecep$
\beq
\mu_{(n_1 ... n_d)}^{(n)} (\vecep) = \sum_{k=0}^{\infty} \frac{1}{k!} \sum_{\substack{k_1... k_d=0 \\ \sum_i k_i=k}}^k \phantom{,} _{(n)}^{(k)}\chi_{(n_1 ... n_d)}^{(k_1 ... k_d)} \, \ecep_1^{k_1} ... \ecep_d^{k_d}
\label{expand}
\eeq
Inserting expansion (\ref{expand}) into the cumulant hierarchy, eq. (\ref{cumul}), and
matching order by order in $k$, we derive another interesting hierarchy for the response coefficients 
of the different cumulants. For $k=0$ this reads
\beq
n_{\alpha} {\cal L}_{\beta \alpha} \phantom{,} _{(n)}^{(0)}\chi_{(n_1 ... n_{\alpha}-1 ... n_{\beta}+1 ... n_d)}^{(0 ... 0)} = 0 \, ,
\label{nonlincoefs1}
\eeq
which is a symmetry relation for the equilibrium ($\vecep=0$) current cumulants. For $k\ge 1$ we obtain
\begin{eqnarray}
\sum_{\substack{k_1... k_d=0 \\ \sum_i k_i=k\ge 1}}^k \Bigg[ \frac{n_{\alpha}}{k} {\cal L}_{\beta \alpha} \phantom{,} 
_{(n)}^{(k)}\chi_{(n_1 ... n_{\alpha}-1 ... n_{\beta}+1 ... n_d)}^{(k_1 ... k_d)} \nonumber \\
+ {\cal L}_{\gamma\nu} \, _{(n+1)}^{(k-1)}\chi_{(n_1 ... n_{\gamma}+1 ... n_d)}^{(k_1... k_{\nu}-1... k_d)}\Bigg] = 0 \, ,
\label{nonlincoefs2}
\end{eqnarray}
which relates $k$-order response coefficients of $n$-order cumulants with $(k-1)$-order coefficients of $(n+1)$-order cumulants. 
Relations (\ref{nonlincoefs1})-(\ref{nonlincoefs2}) for the response coefficients result from the IFR in the limit of infinitesimal rotations. 
For a finite rotation ${\cal R}=-\mathbb{I}$, which is equivalent to a current inversion, we have $\mu(\vla)=\mu(-\vla-2\vecep)$ and we may use this in the definition of response coefficients, 
$_{(n)}^{(k)}\chi_{(n_1 ... n_d)}^{(k_1 ... k_d)} \equiv k! [\partial^{n+k} \mu(\vla)/ \partial \lambda_1^{n_1}...\lambda_d^{n_d} \partial \ecep_1^{k_1}...\ecep_d^{k_d}]_{\lambda=0=\ecep}$, 
see eq. (\ref{expand}), to obtain a complementary relation for the response coefficients
\beq
_{(n)}^{(k)}\chi_{(n_1 ... n_d)}^{(k_1 ... k_d)} = k! \sum_{p_1=0}^{k_1}... \sum_{p_d=0}^{k_d} \frac{(-1)^{n+p} 2^p}{(k-p)!} \phantom{,} _{(n+p)}^{(k-p)}\chi_{(n_1+p_1 ... n_d+p_d)}^{(k_1-p_1 ... k_d-p_d)} \, ,
\label{nonlincoefs3}
\eeq
where $p=\sum_i p_i$. A similar equation was derived in \cite{Gaspard} from the standard fluctuation theorem, although the IFR adds further relations. 
All together, eqs. (\ref{nonlincoefs1})-(\ref{nonlincoefs3}) imply deep relations between the response coefficients at arbitrary orders which go far beyond Onsager's reciprocity relations and Green-Kubo formulae.
As an example, we discuss in the main text some of these relations for a two-dimensional system.


\appendix[Hydrodynamic fluctuation theory]

\label{ap1}
The evolution of the system of interest is described by the following Langevin equation
\beq
\partial_t \rho(\vrr,t) = -\vnabla \cdot \Big( \vQ_{\vE}[\rho(\vrr,t)] + \vxi(\vrr,t) \Big) \, ,
\label{Alangevin}
\eeq
which expresses the local conservation of certain physical observable.
Here $\rho(\vrr,t)$ is the density field, $\vj (\vrr,t)\equiv \vQ_{\vE}[\rho(\vrr,t)] + \vxi(\vrr,t)$ is the fluctuating current, 
with local average $\vQ_{\vE}[\rho(\vrr,t)]$, and $\vxi(\vrr,t)$ is a Gaussian white noise with zero mean and characterized by a 
variance (or mobility) $\sigma[\rho(\vrr,t)]$.
Notice that the current functional includes in general the effect of a conservative external field, $\vQ_{\vE}[\rho(\vrr,t)]=\vQ[\rho(\vrr,t)] + \sigma[\rho(\vrr,t)] \vE$.
Using a path integral formulation \cite{Carlos}, the probability of observing a given history $\{\rho(\vrr,t),\vj(\vrr,t)\}_0^{\tau}$ of duration $\tau$ 
for the density and current fields can be written as
\beq
\text{P}\left(\{\rho,\vj\}_0^{\tau} \right) \sim \exp \Big( +L^d I_{\tau}\left[\rho,\vj \right] \Big) \, ,
\label{AHFT}
\eeq
where $L$ is the system linear size, $d$ is the dimensionality, and the functional $I_{\tau}\left[\rho,\vj \right]$ is
\beq
I_{\tau}\left[\rho,\vj \right]= - \int_0^{\tau} dt \int d \vrr \frac{\displaystyle \Big(\vj(\vrr,t) - \vQ_{\vE}[\rho(\vrr,t)] \Big)^2}
{\displaystyle 2 \sigma[\rho(\vrr,t)] } \,  ,
\label{AHFTfunc}
\eeq
with $\rho (\vrr,t)$ and $\vj(\vrr,t)$ coupled via the continuity equation 
\beq
\partial_t \rho (\vrr,t) + \vnabla\cdot \vj(\vrr,t) = 0 \, . 
\label{Aconteq}
\eeq
In this way the probability of each history $\{\rho,\vj\}_0^{\tau}$ has a Gaussian weight around the average local behavior given by $\vQ_{\vE}[\rho(\vrr,t)]$. 
Eqs. (\ref{AHFT}) and (\ref{AHFTfunc}) are equivalent to the hydrodynamic fluctuation theory recently proposed by Bertini and coworkers \cite{BertiniA,BertiniB,BertiniC}.
The probability $\text{P}_{\tau}(\vJJ)$  of observing a space- and time-averaged empirical current $\vJJ$, defined as
\beq
\vJJ = \frac{1}{\tau}  \int_0^{\tau} dt \int d\vrr \, \, \vj (\vrr,t) \, ,
\label{AempJ}
\eeq
can be obtained from the path integral of $\text{P}\left(\{\rho,\vj\}_0^{\tau} \right)$ restricted to histories $\{\rho,\vj\}_0^{\tau}$ compatible with a given $\vJJ$,
\beq
\text{P}_{\tau}(\vJJ) = \int {\cal D} \rho \, {\cal D} \vj \, \text{P}\left(\{\rho,\vj\}_0^{\tau} \right) \delta \Big(\vJJ - 
\frac{1}{\tau}  \int_0^{\tau} dt \int d\vrr \, \, \vj (\vrr,t)  \Big) \, ,
\label{pathint}
\eeq
This probability scales for long times as $\text{P}_{\tau}(\vJJ)\sim \exp[+\tau L^d G(\vJJ)]$, and the current large deviation function (LDF) $G(\vJJ)$ can be 
related to $I_{\tau}[\rho,\vj]$ via a simple saddle-point calculation in the long-time limit,
\beq
G(\vJJ) = \lim_{\tau \to \infty} \frac{1}{\tau} \max_{\rho(\vrr,t) \atop \vj(\vrr,t)} I_{\tau}[\rho,\vj] \, ,
\label{ldf0}
\eeq
subject to constraints (\ref{Aconteq}) and (\ref{AempJ}). The density and current fields solution of this variational problem, denoted here as $\rho_0(\vrr,t;\vJJ)$ and $\vj_0(\vrr,t;\vJJ)$, can be interpreted as 
the optimal path the system follows in order to sustain a lont-time current fluctuation $\vJJ$. It is worth emphasizing here that the existence of an optimal path
rests on the presence of a selection principle at play, namely a long time, large size limit which selects, among all possible paths compatible with a given fluctuation, an optimal
one via a saddle point mechanism.
Eq. (\ref{ldf0}) defines a complex spatiotemporal problem whose solution remains challenging in most cases \cite{Carlos,BertiniA,BertiniB,BertiniC,BD,PabloA,PabloB,BD2,Pablo3}. 
However, the following hypotheses greatly reduce its complexity:
\begin{enumerate}
\item[(i)] We assume that the optimal profiles responsible of a given current fluctuation are time-independent \cite{BD}, $\rho_0(\vrr;\vJJ)$ and 
$\vj_0(\vrr;\vJJ)$. This, together with the continuity equation, implies that the optimal current vector field is divergence-free, 
$\vnabla\cdot\vj_0(\vrr;\vJJ)=0$.
\item[(ii)] A further simplification consists in assuming that the optimal current field is in fact constant across space, so $\vj_0(\vrr;\vJJ)=\vJJ$.
\end{enumerate}
Provided that these hypotheses hold, the current LDF can be written as
\beq
G(\vJJ) =-\min_{\rho(\vrr)} \int \frac{\displaystyle \left(\vJJ - \vQ_{\vE}[\rho(\vrr)]\right)^2} {\displaystyle 2 \sigma[\rho(\vrr)] } d \vrr \, ,
\label{Aldf}
\eeq
The optimal density profile is thus solution of the following differential equation
\beq
\frac{\delta \om_2[\rho(\vrr)]}{\delta \rho(\vrr')} - 2\vJJ\cdot \frac{\delta \vom_1[\rho(\vrr)]}{\delta \rho(\vrr')} + 
\vJJ^2 \frac{\delta \om_0[\rho(\vrr)]}{\delta \rho(\vrr')}  = 0 \, ,
\label{Aoptprof}
\eeq
where $\frac{\delta}{\delta \rho(\vrr')}$ stands for functional derivative, and
\beq
\vom_n[\rho(\vrr)] \equiv  \int d\vrr \, \textbf{W}_n[\rho(\vrr)] \qquad \text{with} \qquad \textbf{W}_n[\rho(\vrr)]\equiv \frac{\vQ_{\vE}^n[\rho(\vrr)]}{\sigma[\rho(\vrr)]} \, .
\label{defs1}
\eeq
For time-reversible systems, one can see that the evolution operator in the Fokker-Planck formulation of
eq. (\ref{Alangevin}) obeys a local detailed balance condition, and
\beq
\textbf{W}_1[\rho(\vrr)] = \frac{\vQ_{\vE}[\rho(\vrr)]}{\sigma[\rho(\vrr)]} = -\vnabla \frac{\delta {\cal H}[\rho]}{\delta\rho} \, ,
\label{detbal}
\eeq
where ${\cal H}[\rho(\vrr)]$ is the system Hamiltonian.
In this case, by using vector integration by parts, it is easy to show that
\beq
\frac{\delta}{\delta \rho(\vrr')} \int d\vrr \textbf{W}_1[\rho(\vrr)] \cdot \cvA(\vrr) = - \frac{\delta}{\delta \rho(\vrr')} \int_{\partial\Lambda} d\Gamma \frac{\delta {\cal H}[\rho]}{\delta \rho} \cvA(\vrr) \cdot \hat{n} =0 \, ,
\label{Acond1}
\eeq
for any divergence-free vector field $\cvA(\vrr)$. The second integral is taken over the boundary $\partial \Lambda$ of the domain $\Lambda$ where the system is defined, 
and $\hat{n}$ is the unit vector normal to the boundary at each point. In particular, by taking $\cvA(\vrr)=\vJJ$ constant, eq. (\ref{Acond1}) implies that $\delta \vom_1[\rho(\vrr)]/\delta \rho(\vrr') = 0$.
Hence for time-reversible systems 
the optimal profile $\rho_0(\vrr;\vJJ)$ remains invariant under rotations of the current $\vJJ$, see eq. (\ref{Aoptprof}),
and this allows us to prove the isometric fluctuation relation (IFR), eqs. (1) and (8) in the main text.

We can now relax hypothesis $\text{(ii)}$ above and study cases where the current profile is not constant. Let $\text{P}_{\tau}[\cvJ(\vrr)]$
be the probability of observing a time-averaged
current field $\cvJ(\vrr)=\tau^{-1}\int_0^{\tau} dt \, \vj (\vrr,t)$. Notice that this vector field must be divergence-free because of hypothesis $(\text{i})$. 
This  probability also obeys a large deviation principle, $\text{P}_{\tau}[\cvJ(\vrr)]\sim \exp\left(+\tau L^d G[\cvJ(\vrr)] \right)$, with a current LDF equivalent to that 
in eq. (\ref{Aldf}) but with a space-dependent current field $\cvJ(\vrr)$. The optimal density profile $\rho_0[\vrr;\cvJ(\vrr)]$ is now solution of
\beq
\frac{\delta}{\delta \rho(\vrr')}\int d\vrr \, \Big(W_2[\rho(\vrr)] - 2 \cvJ(\vrr) \cdot \textbf{W}_1[\rho(\vrr)] +  \cvJ^2(\vrr)W_0[\rho(\vrr)] \Big)   = 0 \, ,
\label{optprofJr}
\eeq
which is the equivalent to eq. (\ref{Aoptprof}) in this case.
For time-reversible systems condition (\ref{Acond1}) holds
and $\rho_0[\vrr;\cvJ(\vrr)]$ remains invariant under (local or global) rotations of $\cvJ(\vrr)$. In this way we can simply relate 
$\text{P}_{\tau}[\cvJ(\vrr)]$ with the probability of any other divergence-free current field $\cvJ'(\vrr)$ locally-isometric to 
$\cvJ(\vrr)$, i.e. $\cvJ'(\vrr)^2 = \cvJ(\vrr)^2$ $\forall \vrr$, via a generalized isometric fluctuation relation, see eq. (12) in the paper.
Notice that in general an arbitrary local or global rotation of a divergence-free vector field does not conserve the zero-divergence property, so this constraints
the current fields and/or local rotations for which this generalized IFR applies.

The large deviation function for the space- and time-averaged current, $G(\vJJ)$, can be related to $G[\cvJ(\vrr)]$ via a contraction principle
\beq
G(\vJJ)=\max_{\substack{\tiny \cvJ(\vrr) : \vnabla\cdot\cvJ(\vrr)=0 \\ \tiny \vJJ=\int d\vrr \, \cvJ(\vrr)}} G[\cvJ(\vrr)] \, .
\label{contrac}
\eeq
The optimal, divergence-free current field $\cvJ_0(\vrr;\vJJ)$ solution of this variational problem may have spatial structure in general.
However, numerical results and phenomenological arguments strongly suggest that the constant solution, $\cvJ_0(\vrr;\vJJ)=\vJJ$, is the optimizer at least 
for a wide interval of current fluctuations, showing that hypothesis (ii) above is not only plausible but also well justified on physical grounds.
In any case, the range of validity of this hypothesis can be explored by studying
the limit of local stability of the constant current solution using tools similar to those in Ref. \cite{BD2}.

Hypotheses (i) and (ii) are the straightforward generalization to $d$-dimensional systems of the Additivity Principle recently conjectured 
by Bodineau and Derrida for one-dimensional diffusive systems \cite{BD}. This conjecture, which has been recently confirmed for a broad current interval in extensive 
simulations of a general diffusion model \cite{PabloA,PabloB}, is however known to break down in some special cases for extreme current 
fluctuations, where time-dependent profiles in the form of traveling waves propagating along the current direction may emerge \cite{BertiniA,BertiniB,BertiniC,BD2,Pablo3}. 
As in previous cases, we can now study the probability $\text{P}\left(\{\vj(\vrr,t)\}_0^{\tau} \right)$ of observing a particular history for 
the current field, which can be written as the path integral of the probability in eq. (\ref{AHFT}) over histories of the density field $\{\rho(\vrr,t)\}_0^{\tau}$ coupled to the desired current field 
via the continuity eq. (\ref{Aconteq}) at every point in space and time. This probability obeys another large deviation principle, 
with an optimal history of the density field $\{\rho_0(\vrr,t)\}_0^{\tau}$ which is solution of an equation similar to eq. (\ref{optprofJr}) but with time-dependent profiles.
However, as opposed to the cases above, the current field $\vj(\vrr,t)$ is not necessarily divergence-free because of the time-dependence of the associated 
$\rho_0(\vrr,t)$, resulting in a violation of condition (\ref{Acond1}). In this way the optimal $\rho_0(\vrr,t)$ depends on both $\vj(\vrr,t)$ and $\vj(\vrr,t)^2$ so it does 
not remain invariant under (local or global) instantaneous rotations of the current field, resulting in a violation of the generalized isometric fluctuation relation in the time-dependent regime.

Notice that the dynamic phase transition to time-dependent optimal paths is expected to occur only for extreme current fluctuations, thus rendering valid the 
isometric fluctuation relations for a wide, subcritical current interval.
Interestingly, we can use the IFR to detect such dynamic phase transition. If we measure $\text{P}_{\tau}[\cvJ(\vrr)]$ in a system described by eq. (\ref{Alangevin}) at the macroscopic level, 
finding that the measured probabilities do not obey the generalized IFR, then we can conclude that such a violation of IFR is due to the onset of time-dependent optimal profiles, thus signaling 
the dynamic phase transition. On the other hand, breakdown of the standard IFR (for space- and time-averaged currents) may signal the onset of space-dependent, 
divergence-free optimal current profiles or the aforementioned dynamic phase transition. In this way, the combined use of the IFR and its generalizations is capable of a full characterization 
of the instabilities which characterize the fluctuating behavior of the system at hand \cite{BD2,Pablo3}.

\appendix[Constants of motion]

\label{ap2}
\noindent A sufficient condition for the IFR to hold is that
\beq
\frac{\delta \vom_1[\rho(\vrr)]}{\delta \rho(\vrr')} = 0 \, ,
\label{cond0}
\eeq
with the functional $\vom_1[\rho(\vrr)]$ defined in eq. (\ref{defs1}) above. We have shown that condition (\ref{cond0}) follows from the time-reversibility of the dynamics, 
in the sense that the evolution operator in the Fokker-Planck formulation of eq. (\ref{langevin}) obeys a local detailed balance condition, see eq. (\ref{Acond1}).
Condition (\ref{cond0}) implies that $\vom_1[\rho(\vrr)]$ is in fact a \emph{constant of motion}, $\vep$, independent of the profile $\rho(\vrr)$. Therefore 
we can use an arbitrary profile $\rho(\vrr)$, compatible with boundary conditions, to compute $\vep$. 
We now choose boundary conditions to be gradient-like in the $\hat{x}$-direction, with densities 
$\rho_L$ and $\rho_R$ at the left and right reservoirs, respectively, and periodic boundary conditions in all other directions.
Given these boundaries, we now select a linear profile 
\beq
\rho(\vrr)=\rho_L + (\rho_R-\rho_L) x \, ,
\label{linprof}
\eeq
to compute $\vep$, with $x\in[0,1]$, and assume very general forms for the current and mobility functionals
\begin{eqnarray}
\vQ[\rho(\vrr)] & \equiv & D_{0,0}[\rho] \vnabla \rho + \sum_{n,m>0} D_{nm}[\rho]  (\vnabla^m \rho)^{2n} \vnabla \rho \, , \nonumber \\
\sigma[\rho(\vrr)] & \equiv & \sigma_{0,0}[\rho] + \sum_{n,m>0} \sigma_{nm}[\rho] (\vnabla^m \rho)^{2n} \, , \nonumber
\end{eqnarray}
where as a convention we denote as $F[\rho]$ a generic functional of the profile but not of its derivatives.
It is now easy to show that $\vep=\varepsilon \hat{x}+\vE$, with
\beq
\varepsilon = \int_{\rho_L}^{\rho_R} d \rho \frac{D_{0,0}(\rho)+\sum_{n>0} D_{n1}(\rho)(\rho_R-\rho_L)^{2n}}
{\sigma_{0,0}(\rho)+\sum_{m>0} \sigma_{m1}(\rho)(\rho_R-\rho_L)^{2m}} \, ,
\label{epsilon}
\eeq
and $\hat{x}$ the unit vector along the gradient direction. In a similar way, if the following condition holds
\beq
\frac{\delta}{\delta \rho(\vrr')} \int \vQ[\rho(\vrr)] d\vrr = 0 \, ,
\label{cond2}
\eeq
together with time-reversibility, eq. (\ref{cond0}), the system can be shown to obey an extended isometric fluctuation relation which links
any current fluctuation $\vJJ$ in the presence of an external field $\vE$ with any other isometric current fluctuation $\vJJ'$ in the presence 
of an arbitrarily-rotated external field $\vE^*$, and reduces to the standard IFR for $\vE=\vE^*$, see eq. (11) in the paper.
Condition (\ref{cond2}) implies that $\vnu\equiv\int \vQ[\rho(\vrr)] d\vrr$ is another constant of motion, which can be now written 
as $\vnu=\nu\hat{x}$, with
\beq
\nu = \int_{\rho_L}^{\rho_R} d \rho \left[D_{0,0}(\rho)+\sum_{n>0} D_{n1}(\rho)(\rho_R-\rho_L)^{2n}\right] \, ,
\label{nu}
\eeq
As an example, for a diffusive system $\vQ[\rho(\vrr)]=-D[\rho] \vnabla \rho(\vrr)$, with $D[\rho]$ the diffusivity functional, and the above equations 
yield the familiar results 
\begin{eqnarray}
\varepsilon & = & \int_{\rho_R}^{\rho_L}  \frac{D(\rho)}{\sigma(\rho)}d\rho \, , \nonumber \\
\phantom{aaa} \nonumber \\
\nu & = & \int_{\rho_R}^{\rho_L} D(\rho)d\rho \, , \nonumber
\end{eqnarray}
for a standard local mobility $\sigma[\rho]$.



\begin{acknowledgments}
Financial support from Spanish MICINN project FIS2009-08451, University of Granada, and Junta de Andaluc\'ia is acknowledged.
\end{acknowledgments}





\end{article}








\end{document}